\documentclass[11pt]{article}

\usepackage[dvips]{graphicx}
\usepackage{epsfig}
\usepackage{version}

\usepackage{geometry}
\usepackage{amssymb,amsmath,eepic,graphics,color}

\usepackage{epsfig}

\newenvironment{proof}{\noindent\textit{Proof: }}{{\hfill $\Box$}}
\newtheorem{lemma}{Lemma}[section]
\newtheorem{theorem}[lemma]{Theorem}
\newtheorem{proposition}[lemma]{Proposition}
\newtheorem{corollary}[lemma]{Corollary}
\newtheorem{reduction}{Rule}

\newtheorem{definition}[lemma]{Definition}

\newtheorem{claim}[lemma]{Claim}

\newtheorem{observation}[lemma]{Observation}

\newcommand{\eg}{\textit{e.g.}}
\newcommand{\gbc}{\textsc{Not-1-in-3-edge-triangle}}
\newcommand{\tgbc}{\textsc{Tripartite-Not-1-in-3-edge-triangle}}
\newcommand{\hyp}{NP \subseteq coNP / poly}

\begin{document}

\title{\textbf{On the (non-)existence of polynomial kernels for $P_l$-free edge modification problems}
\thanks{Research supported by the AGAPE project  (ANR-09-BLAN-0159).}}

\author{Sylvain Guillemot$^1$ \and Christophe Paul$^2$ \and Anthony Perez$^2$ \\\\
$^1$ Lehrstuhl f\"ur Bioinformatik, Friedrich-Schiller Universit\"at Jena \\
$^2$ Universit\'e Montpellier II - CNRS, LIRMM}

\date{}

\maketitle

\begin{abstract}
Given a graph $G=(V,E)$ and an integer $k$, an edge modification problem for a graph property $\Pi$ consists in deciding whether there exists a set of edges $F$ of size at most $k$ such that the graph $H=(V,E\vartriangle F)$ satisfies the property $\Pi$. In the $\Pi$ \emph{edge-completion problem}, the set $F$ of edges is constrained to be disjoint from $E$; in the $\Pi$ \emph{edge-deletion problem}, $F$ is a subset of $E$; no constraint is imposed on $F$ in the $\Pi$ \emph{edge-edition problem}. 
A number of optimization problems can be expressed in terms of graph modification problems which have been extensively studied in the context of parameterized complexity. When parameterized by the size $k$ of the edge set $F$, it has been proved that if $\Pi$ is an hereditary property characterized by a finite set of forbidden induced subgraphs, then the three $\Pi$ edge-modification problems are FPT~\cite{Cai96}. It was then natural to ask~\cite{Cai96} whether these problems also admit a polynomial size kernel. 
Using recent lower bound techniques, Kratsch and Wahlstr\"om answered this question negatively~\cite{KW09}. However, the problem remains open on many natural graph classes characterized by forbidden induced subgraphs. 
Kratsch and Wahlstr\"om asked whether the result holds when the forbidden subgraphs are paths or cycles and pointed out that the problem is already open in the case of $P_4$-free graphs (i.e. cographs). This paper provides positive and negative results in that line of research. We prove that parameterized cograph edge modification problems have cubic vertex kernels whereas polynomial kernels are unlikely to exist for the $P_l$-free and $C_l$-free edge-deletion problems for large enough $l$.
\end{abstract}

\section{Introduction}

An edge modification problem aims at changing the edge set of an input graph $G=(V,E)$ in order to get a certain property $\Pi$ satisfied (see~\cite{NSS01} for a recent study). Edge modification problems cover a broad range of graph optimization problems among which completion problems (\eg ~\textsc{minimum fill-in}, \emph{a.k.a}  \textsc{chordal graph completion}~\cite{Ros72,TY84}), edition problems (\eg ~\textsc{cluster editing}~\cite{SST04}) and edge deletion problems (\eg ~\textsc{maximum planar subgraph}~\cite{GJ79}).  In a completion problem, the set $F$ of modified edges is constrained to be disjoint from $E$; in an edge deletion problem, $F$ has to be a subset of $E$; and in an edition problem, no restriction applies to $F$. These problems are fundamental in graph theory and play an important role in computational complexity theory (indeed they represent a large number of the earliest NP-Complete problems~\cite{GJ79}). Edge modification problems are also relevant in the context of applications as graphs are often used to model data sets which may contain errors. Adding or deleting an edge thereby corresponds to fixing some false negatives or false positives (see \emph{e.g.}~\cite{SST04} in the context of \textsc{cluster editing}). Different variants of edge modification problems have been studied in the literature such as graph sandwich problems~\cite{GKS95}.
Most of the edge modification problems turns out to be NP-Complete~\cite{NSS01} and approximation algorithms exist for some known graph properties (see \eg~\cite{KS07,ZW08}). But for those who want to compute an exact solution, fixed parameter algorithms~\cite{DF99,FG06,Nie06} are a good alternative to cope with such hard problems. In the last decades, edge modification problems have been extensively studied in the context of fixed parameterized complexity (see~\cite{Cai96,FLR07,HPTV07}). 

A parameterized problem $Q$ is \emph{fixed parameter tractable} (FPT for short) with respect to parameter $k$ whenever it can be solved in time $f(k).n^{O(1)}$, where $f(k)$ is an arbitrary computable function~\cite{DF99,Nie06}. 
In the context of edge modification problems, the size $k$ of the set $F$ of modified edges is a natural parameterization. The generic question is thereby whether a given edge modification problem is FPT for this parameterization. 
More formally:

\medskip
\noindent
\textsc{Parameterized $\Pi$ edge--modification Problem}\\
\textbf{Input:} An undirected graph $G=(V,E)$.\\
\textbf{Parameter:} An integer $k\geqslant 0$.\\
\textbf{Question:} Is there a subset $F\subseteq V\times V$ with
$|F|\leqslant k$ such that the graph $H=(V,E\vartriangle F)$ satisfies $\Pi$.

\medskip
A classical result of parameterized complexity states that a parameterized problem $Q$ is FPT if and only if it admits a \emph{kernelization}. 
A \emph{kernelization} of a parameterized problem $Q$ is a polynomial time algorithm $\mathcal{K}$ that given an instance $(x,k)$ computes an equivalent instance $\mathcal{K}(x,k)=(x',k')$ such that the size of $x'$ and $k'$ are bounded by a computable function $h()$ depending only on the parameter $k$. The reduced instance $(x',k')$ is called a \emph{kernel} and we say that $Q$ admits a \emph{polynomial kernel} if the function $h()$ is a polynomial. The equivalence between  the existence of an FPT algorithm and the existence of a kernelization only yields kernels of (at least) exponential size. Determining whether an FPT problem has a polynomial (or even linear) size kernel is thus an important challenge. Indeed, the existence of such polynomial time reduction algorithm (or pre-processing algorithm or \emph{reduction rules}) really speed-up the resolution of the problem, especially if it is interleaved with other techniques~\cite{NR00}. However, recent results proved that not every fixed parameter tractable problem admits a polynomial kernel~\cite{BDFH08}. 

Cai~\cite{Cai96} proved that if $\Pi$ is an hereditary graph property characterized by a finite set of forbidden subgraphs, then the \textsc{parameterized $\Pi$ modification} problems (edge-completion, edge-deletion and edge-edition) are FPT. It was then natural to ask~\cite{Cai96} whether these $\Pi$ edge-modification problems also admit a polynomial size kernel. Using recent lower bound techniques, Kratsch and Wahlstr\"om answered negatively this question~\cite{KW09}. However, the problem remains open on many natural graph classes characterized by forbidden induced subgraphs. 
Kratsch and Wahlstr\"om asked whether the result holds when the forbidden subgraphs are paths or cycles and pointed out that the problem is already open in the case of $P_4$-free graphs (i.e. cographs). In this paper, we prove that \textsc{parameterized cograph edge modification} problems have cubic vertex kernels whereas polynomial kernels are unlikely to exist for \textsc{$P_l$-free} and \textsc{$C_l$-free edge deletion} problems for large enough $l$.
The NP-Completeness of the cograph edge-deletion and edge-completion problems have been proved in~\cite{EC88}.

\paragraph{Outline of the paper.}
We first establish structural properties of optimal edge-modification sets with respect to modules of the input graph (Section 2). These properties allow us to design general reduction rules (Section 3.1). We then establish cubic kernels using an extra sunflower rule (Section 3.2 and 3.3). Finally, we show it is unlikely that the \textsc{$C_l$-free} and the \textsc{$P_l$-free edge-deletion} problems have polynomial kernels (Section 4).


\section{Preliminaries}

\subsection{Notations} 

We only consider finite undirected graphs without loops nor multiple edges. Given a graph $G=(V,E)$, we denote by $xy$ the edge of $E$ between the vertices $x$ and $y$ of $V$. We set $n=|V|$ and $m=|E|$ (subscripts will be used to avoid possible confusion). The neighbourhood of a vertex $x$ is denoted by $N(x)$. If $S$ is a subset of vertices, then $G[S]$ is the subgraph induced by $S$ (i.e. any edge $xy\in E$ between vertices $x,y\in S$ belongs to $E_{G[S]}$). Given a set of pairs of vertices $F$ and a subset $S \subseteq V$, $F[S]$ denotes the pairs of $F$ with both vertices in $S$. 
Given two sets $S$ and $S'$, we denote by $S\vartriangle S'$ their symmetric difference.

\subsection{Fixed parameter complexity and kernelization}

We let $\Sigma$ denote a finite alphabet and $\mathbb{N}$ the set of natural numbers. A \emph{(classical) problem} $Q$ is a subset of $\Sigma^*$, and a string $x\in \Sigma^*$ is an \emph{input} of $Q$. A \emph{parameterized problem} $Q$ over $\Sigma$ is a subset of $\Sigma^*\times \mathbb{N}$. The second component of an input $(x,k)$ of a parameterized problem is called the \emph{parameter}. Given a parameterized problem $Q$, one can derive its unparameterized (or classical) version $\tilde{Q}$ by $\tilde{Q}=\{x\#1^k:(x,k)\in Q\}$, where $\#$ is a symbol that does not belong to $\Sigma$.

A parameterized problem $Q$  is \emph{fixed parameter tractable} (FPT for short) if there is an algorithm which given an instance $(x,k)\in \Sigma^*\times \mathbb{N}$ decides whether $(x,k)\in Q$
in time $f(k).n^{O(1)}$ where $f(k)$ is an arbitrary computable function (see~\cite{DF99,FG06,Nie06}). 
A \emph{kernelization} of a parameterized problem $Q$ is a polynomial time algorithm $\mathcal{K}:
\Sigma^*\times \mathbb{N}\rightarrow \Sigma^*\times \mathbb{N}$ which given an instance $(x,k)\in \Sigma^*\times \mathbb{N}$ outputs an instance $(x',k')\in \Sigma^*\times \mathbb{N}$ such that 
\begin{enumerate}
\item $(x,k)\in Q \Leftrightarrow (x',k')\in Q$ and

\item $|x'|,k'\leqslant h(k)$ for some computable function $h: \mathbb{N}\rightarrow \mathbb{N}$.
\end{enumerate}

The reduced instance $(x',k')$ is called a \emph{kernel} and we say that $Q$ admits a \emph{polynomial kernel} if the function $h()$ is a polynomial. 
It is well known that a parameterized problem $Q$ is FPT if and only if it has a kernelization~\cite{Nie06}. But this equivalence only yields (at least) exponential size kernels. Recent results proved that it is unlikely that every fixed parameter tractable problem admits a polynomial kernel~\cite{BDFH08}. These results rely on the notion of \emph{(or-)composition algorithms} for parameterized problems, which together with a polynomial kernel would imply a collapse on the polynomial hierarchy~\cite{BDFH08}.
An \emph{or-composition algorithm} for a parameterized problem $Q$ is an algorithm that receives as input a sequence of instances $(x_1,k)\dots (x_t,k)$ with $(x_i,k)\in \Sigma^*\times\mathbb{N}$ for $1\leqslant i\leqslant t$, runs in time polynomial in $\sum_{i=1}^t |x_i|+k$ 
and outputs an instance $(y,k')$ of $Q$ such that:
\begin{enumerate}
\item $(y,k')\in Q \Leftrightarrow (x_i,k)\in Q$ for some $1\leqslant i\leqslant t$ and
\item $k'$ is polynomial in $k$.
\end{enumerate}

A parameterized problem admitting an \emph{or-composition algorithm} is said to be \emph{or-compositional}.

\begin{theorem}\cite{BDFH08,FS08}\label{th:or-comp}
Let $Q$ be an or-compositional parameterized problem whose unparameterized version $\tilde{Q}$ is NP-complete. The problem $Q$ does not admit a polynomial kernel unless $NP\subseteq coNP/Poly$.
\end{theorem}

Let $P$ and $Q$ be parameterized problems. A \emph{polynomial time and parameter transformation} from $P$ to $Q$ is a polynomial time computable function $\mathcal{T}:\Sigma^*\times \mathbb{N}\rightarrow \Sigma^*\times \mathbb{N}$ which given an instance $(x,k)\in \Sigma^*\times \mathbb{N}$ outputs an instance $(x',k')\in \Sigma^*\times \mathbb{N}$ such that 
\begin{enumerate}
\item $(x,k)\in P \Leftrightarrow (x',k')\in Q$ and

\item $k'\leqslant p(k)$ for some polynomial $p$.
\end{enumerate}

\begin{theorem}\cite{BTY09}\label{th:ppt}
Let $P$ and $Q$ be parameterized problems and let $\tilde{P}$ and $\tilde{Q}$ be their unparameterized versions. Suppose that $\tilde{P}$ is NP-complete and $\tilde{Q}$ belongs to NP. If there is a polynomial time and parameter transformation from $P$ to $Q$ and if $Q$ admits a polynomial kernel, then $P$ also admits a polynomial kernel.
\end{theorem}

\subsection{Modular decomposition and cographs}

A \emph{module} in a graph $G=(V,E)$ is a set of vertices $M\subseteq V$ such that for any $x\notin M$ either $M\subseteq N(x)$ or $M\cap N(x)=\emptyset$. Clearly if $M=V$ or $|M|=1$, then $M$ is a \emph{trivial} module. 
A graph without any non-trivial module is called \emph{prime}. 
For two disjoint modules $M$ and $M'$, either all the vertices of $M$ are adjacent to all the vertices of $M'$ or none of the vertices of $M$ is adjacent to any vertex of $M'$. 
A partition $\mathcal{P}=\{M_1,\dots M_k\}$ of the vertex set $V(G)$ whose parts are modules is a \emph{modular partition}. A \emph{quotient graph} $G_{/\mathcal{P}}$ is associated with any modular partition $\mathcal{P}$: its vertices are the parts of $\mathcal{P}$ and there is an edge between $M_i$ and $M_j$ iff $M_i$ and $M_j$ are adjacent in $G$.

\begin{figure}[t]
\centerline{\includegraphics[scale=0.75]{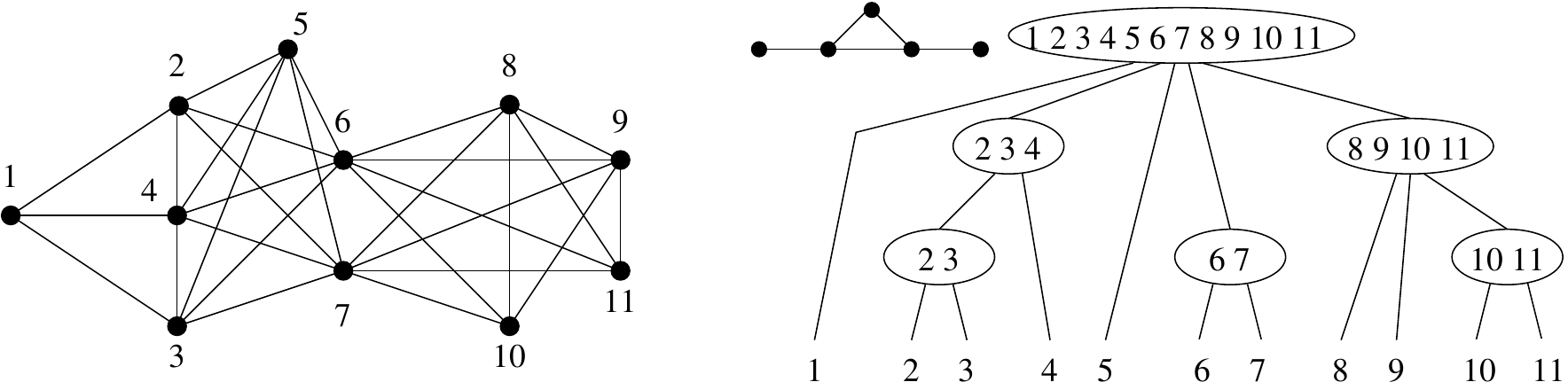}}
\caption{A graph $G$ and its modular decomposition tree $MD(G)$. The root of $MD(G)$ is prime and its quotient graph is the 5 vertex graph depicted eside. Every other node is either parallel or series.
\label{fig:md-tree}
}
\end{figure}

A module $M$ is \emph{strong} if for any module $M'$ distinct from $M$, either $M\cap M'=\emptyset$ or $M\subset M'$ or $M'\subset M$. It is clear from definition that the family of strong modules arranges in an inclusion tree, called the \emph{modular decomposition tree} and denoted $MD(G)$. Each node $N$ of $MD(G)$ is associated with a quotient graph $G_N$ whose vertices correspond to the children $N_1,\dots N_k$ of $N$. (see Figure~\ref{fig:md-tree} for an example). 
We say that a node $N$ of $MD(G)$ is \emph{parallel} if $G_N$ has no edge, \emph{series} if $G_N$ is complete, and \emph{prime} otherwise. 
For a survey on modular decomposition theory, refer to~\cite{HP10}.

\begin{definition}
Let  $G_1=(V_1,E_1)$ and $G_2=(V_2,E_2)$ be two vertex disjoint graphs. The \emph{series composition} of $G_1$ and $G_2$ is the graph $G_1\otimes G_2=(V_1\cup V_2,E_1\cup E_2\cup V_1\times V_2)$. The \emph{parallel composition} of $G_1$ and $G_2$ is the graph $G_1\oplus G_2=(V_1\cup V_2,E_1\cup E_2)$
\end{definition}

Parallel and series nodes in the modular decomposition tree respectively correspond to a parallel and series composition of their children.  \\

Cographs are commonly known as $P_4$-free graphs (a $P_4$ is an induced path on four vertices). However, they were originally defined as follows:

\begin{definition}[\cite{BLS99}]
\label{def:cographs}
	A graph is a cograph if it can be constructed from single vertex graphs by a sequence of parallel and series composition.
\end{definition}

In particular, this means that the modular decomposition tree of a \emph{cograph} does not contain any prime node. It follows that cographs are also known as the totally decomposable graphs for the modular decomposition.

\section{Polynomial kernels for cograph modification problems}
\label{sec:kernels}

\subsection{Modular decomposition based reduction rules}

Since cographs correspond to $P_4$-free graphs, cograph edge-modification problems consist in adding or deleting at most $k$ edges to the input graph in order to make it $P_4$-free. The use of the modular decomposition tree in our algorithms follows from the following observation: 

\begin{observation}\label{obs:p4} [Folklore]
Let $M$ be a module of a graph $G=(V,E)$ and $\{a,b,c,d\}$ be four vertices inducing a $P_4$ of $G$, then $|M\cap\{a,b,c,d\}|\leqslant 1~\mbox{or}~\{a,b,c,d\}\subseteq M$.
\end{observation}

This means that given a modular partition $\mathcal{P}$ of a graph $G$, any induced $P_4$ of $G$ is either contained in a part of $\mathcal{P}$ or intersects the parts of $\mathcal{P}$ in at most one vertex. This observation allows us to show that a cograph edge-modification problem can be solved independently on modules of the partition $\mathcal{P}$ and on the quotient graph $G_{/\mathcal{P}}$, as stated by the following results:

\begin{observation} \label{obs:module}
Let $M$ be a non-trivial module of a graph $G=(V,E)$. Let $F_M$ be an optimal edge-deletion (resp. edge-completion, edge-edition) set of $G[M]$ and  let $F_{opt}$ be an optimal edge-deletion (resp. edge-completion, edge-edition) set of $G$. Then 
$$F=(F_{opt}\setminus F_{opt}[M])\cup F_M$$
is an optimal edge-deletion (resp. edge-completion, edge-edition) set of $G$.
\end{observation}

\begin{proof}
By Observation~\ref{obs:p4}, it follows that $H=(V,E\vartriangle F)$ is $P_4$-free, thereby $F$ is an edge-deletion set. 
As being a cograph is an hereditary property, $F_{opt}[M]$ is an edge-deletion set of $G[M]$. Now observe that $|F| = |F_{opt}|$  since otherwise $|F_M|>|F_{opt}[M]|$, which would contradict the optimality of $F_M$. The same argument holds for edge-completion and edge-edition sets.
 \end{proof}

\begin{lemma} \label{lem:one-module}
Let $M$  be a module of a graph $G=(V,E)$. There exists an optimal edge-deletion (resp. edge-completion, edge-edition) set $F$ such that $M$ is a module of the cograph $H=(V,E\vartriangle F)$.
\end{lemma}

\begin{proof}
Let $F_{opt}$ be an optimal edge-deletion set  and denote $H_{opt}=(V,E\vartriangle F_{opt})$. Let $x$ be a vertex of $M$ such that $|\{xy\in F:y\notin M\}|$ is minimum.
We argue that the following set of edges is an optimal edge-deletion set:
$$F=F_{opt}[M]\cup F_{opt}[V\setminus M]\cup \{zy: z\in M,y\notin M,xy\in F_{opt}\}$$
First observe that by construction $M$ is a module in the graph $H=(V,E\vartriangle F)$ and that by the choice of $x$, $|F|\leqslant |F_{opt}|$. Let us prove that $H$ is $P_4$-free. As $H[M]$ and $H[V\setminus M]$ are respectively isomorphic to $H_{opt}[M]$ and $H_{opt}[V\setminus M]$, they are $P_4$-free. So if $H$ contains an induced $P_4$, its vertices $\{a,b,c,d\}$ intersect $M$ and $V\setminus M$. As $M$ is a module of $H$ it follows by Observation~\ref{obs:p4} that $|M\cap\{a,b,c,d\}|=1$ (say $a\in M\cap\{a,b,c,d\}$). It follows by construction of $F$, that $\{x,b,c,d\}$ also induces a $P_4$ in $H_{opt}$, contradicting the assumption that $F_{opt}$ is an edge-deletion set. So we proved that $F$ is an edge-deletion set of $G$ which preserves the module $M$ and is not larger than $F_{opt}$. 
The same proof holds for edge-completion and edge-edition sets.
\end{proof}

\begin{lemma}
\label{lem:modules}
Let $G=(V,E)$ be an arbitrary graph. There exists an optimal edge-deletion (resp. edge-completion, edge-edition) set $F$ such that every module $M$ of $G$ is module of the cograph $H=(V,E\vartriangle F)$.
\end{lemma}

\begin{proof}
We prove the statement for edge-deletion sets by induction on the number of modules of a graph. The same proof applies for edge-completion and edge-edition sets. Observe that the result trivially holds if $G$ is a prime graph and follows from Lemma~\ref{lem:one-module} if $G$ contains a unique non-trivial module.  

Let us now assume that the property holds for every graph with at most $t$ non-trivial modules. Let $G$ be a graph with $t+1$ non-trivial modules and let $M$ be a non-trivial module of $G$ which is minimal for inclusion. By induction hypothesis, the statement holds on $G[M]$ (since it is prime) and on the graph $G_{M\rightarrow x}$ where $M$ has been contracted to a single vertex $x$ (since it contains at most $t$ non-trivial modules). The conclusion follows from Observation~\ref{obs:module}.
 \end{proof}\\


We now present three reduction rules which apply to the three cograph edge-modification problems we consider. The second reduction rule is not required to obtain a polynomial kernel for each of these problems. However, it will ease the analysis of the structure of a reduced graph.

\begin{reduction} \label{rule1}
Remove the connected components of $G$ which are cographs.
\end{reduction}

\begin{reduction} \label{rule2}
If $C=G_1\otimes G_2$ is a connected component of $G$, then replace $C$ by $G_1\oplus G_2$.
\end{reduction}

\begin{reduction} \label{rule3}
If $M$ is a non-trivial module of $G$ which is strictly contained in a connected component and is not an independent set of size at most $k+1$, then return the graph $G'\oplus G[M]$ where $G'$ is obtained from $G$ by replacing $M$ by an independent set module of size $min\{|M|,k+1\}$.
\end{reduction}

Observe that if $G[M]$ is a cograph, adding a disjoint copy to the graph is useless since it will then  be removed by Rule~\ref{rule1}. 

\begin{lemma} \label{lem:rules}
The reduction rules \ref{rule1}, \ref{rule2} and \ref{rule3} are safe and can be carried out in linear time.
\end{lemma}

\begin{proof}
The three rules can be computed in linear time using any linear time modular decomposition algorithm~\cite{HP10}. 
The first rule is trivially safe. The second rule is safe by Lemma~\ref{lem:modules}. The safeness of Rule~\ref{rule3} also follows from Lemma~\ref{lem:modules}: there always exists an optimal solution that updates all or none of the edges between any two disjoint modules. Thereby if a module $M$ has size larger than $k+1$, none of the edges (or non-edges) $xy$ with $x\in M$, $y\notin M$ can be changed in such a solution. Shrinking $M$ into an independent set of size $k+1$ and adding a disjoint copy of $G[M]$ (to keep track of the edge modification inside the module) is thereby safe.
\end{proof}\\

The analysis of the size of the kernel relies on the following structural property of the modular decomposition tree of an instance reduced under Rule~\ref{rule1}, Rule~\ref{rule2} and Rule~\ref{rule3}.

\begin{observation} \label{obs:reduced-123}
Let $G$ be a graph reduced under Rule~\ref{rule1}, Rule~\ref{rule2} and Rule~\ref{rule3}. If $C$ is a non prime connected component of $G$, then the modules of $C$ are independent sets of size at most $k+1$.
\end{observation}

\begin{proof}
By Rule~\ref{rule2}, none of the connected components of $G$ results from a series composition. By Rule~\ref{rule3}, a module which is not the union of some connected components of $G$ has size at most $k+1$ and is an independent set.
\end{proof}\\

Observe that these three reduction rules preserve the parameter. However, Rule~\ref{rule3} increases the number of vertices of the instance. Nevertheless, we will be able to bound the number of vertices of a reduced instance. \\

It remains to show that computing a reduced graph requires polynomial time. Let us mention that it is safe to apply Rule~\ref{rule2} and Rule~\ref{rule3} only on strong modules (in Rule~\ref{rule2}, $G_1$ can be chosen as a strong module).  

\begin{lemma} \label{lem:gen-reduced}
Given a graph $G=(V,E)$, computing a graph reduced under Rule~\ref{rule1}, Rule~\ref{rule2} and Rule~\ref{rule3} requires polynomial time.
\end{lemma}

\begin{proof}
Let us say that a module $M$ of $G$ is \emph{reduced} if it is an independent set of size at most $k+1$ or the disjoint union of some connected components of $G$ (observe that connected components of $G$ are also modules of $G$). By Observation~\ref{obs:reduced-123}, if $G$ is reduced under Rule~\ref{rule1}, Rule~\ref{rule2} and Rule~\ref{rule3}, then every module of $G$ is reduced. Notice that if every strong module of $G$ is reduced, then every module of $G$ is reduced.
So to prove the statement, we count the number of strong modules (\textit{i.e.} nodes of the modular decomposition tree $MD(G)$) which are not reduced. 

Let us also remark that if a connected component $C$ is a cograph with at least two vertices, then a series of applications of Rule~\ref{rule2} eventually transforms $C$ in a set of isolated vertices. This means that we can assume that the applications of Rule~\ref{rule1} is postponed to the end of the reduction process. This will ease the argument below.

When Rule~\ref{rule3} is applied, then by definition the number of non-reduced strong modules decreases by one. When Rule~\ref{rule2} is applied, unless $G_1$ is an independent set of size at most $k+1$, then the number of non-reduced strong modules also decreases by one. But observe that if $G_1$ is an independent set of size at most $k+1$, then its vertices will be removed by Rule~\ref{rule1} as they will become isolated vertices. As the number of strong modules of a graph is bounded by the number of vertices, this proves that a series of at most $n$ applications of Rule~\ref{rule2} and Rule~\ref{rule3} is enough to compute a reduced graph.
\end{proof}

\subsection{Cograph edge-deletion (and edge-completion)}

In addition to the previous reduction rules, we need the classical \emph{sunflower} rule to obtain a polynomial kernel for the parameterized cograph edge-deletion problem.

\begin{reduction} \label{rule:sunflower-del}
If $e$ is an edge of $G$ that belongs to a set $\mathcal{P}$ of at least $k+1$ $P_4$'s such that $e$ is the only common edge of any two distinct $P_4$'s of $\mathcal{P}$, then remove $e$ and decrease $k$ by one.
\end{reduction}

\begin{observation} \label{obs:rule-del}
The reduction rule~\ref{rule:sunflower-del} is safe and can be carried out in polynomial time.
\end{observation}

\begin{proof}
It is clear that the edge $e$ has to be deleted as otherwise at least $k+1$ edge deletions would be required to break all the $P_4$'s of the set $\mathcal{P}$. Such an edge, if it exists, can be found in polynomial time if one computes the set of all $P_4$'s of the input graph (which can be done in $O(n^4)$ time).
\end{proof}\\

To analyse the size of a reduced graph $G=(V,E)$, we study the structure of the cograph $H=(V,E\vartriangle F)$ resulting from the removal of an optimal (of size at most $k$) edge-deletion set $F$. The modular decomposition tree (or cotree) is the appropriate tool for this analysis.

\newpage 

\begin{theorem} \label{th:deletion}
The parameterized cograph edge-deletion problem admits a cubic vertex kernel.
\end{theorem}
\begin{proof}
Let $G = (V,E)$ be a graph reduced under Rule~\ref{rule1}, Rule~\ref{rule2}, Rule~\ref{rule3} and Rule~\ref{rule:sunflower-del} that can be turned into a cograph by deleting at most $k$ edges. Let $F$ be an optimal edge-deletion set and denote by $H = (V, E \vartriangle F)$ the cograph resulting from the deletion of $F$ and by $T$ its cotree. We will count the number of leaves of $T$ (or equivalently of vertices of $G$ and $H$).

Observe that since a set of $k$ edges covers at most $2k$ vertices, $T$ contains at most $2k$ affected leaves (i.e. leaves corresponding to a vertex incident to a removed edge).
We say that an internal node of the cotree $T$ is \emph{affected} if it is the least common ancestor of two affected leaves. Notice that there are at most $2k$ affected nodes.

We first argue that the root of $T$ is a parallel node and is affected. Assume that the root of $T$ is a series node: since no edges are added to $G$, this would imply that $G$ is not reduced under Rule~\ref{rule2}, a contradiction. Moreover, since $G$ is reduced under Rule~\ref{rule1}, none of its connected components is a cograph. It follows that every connected component of $G$ contains a vertex incident to a removed edge, and thus that every subtree attached to the root contains an affected leaf as a descendant. Hence the root of $T$ is an affected node.

\begin{claim}
\label{claim:kernel-del}
Let $p$ be an affected leaf or an affected node different from the root, and $q$ be the least affected ancestor of $p$. The path between $p$ and $q$ has length at most $2k + 3$.
\end{claim}

\emph{Proof.} Observe first that the result trivially holds if $q$ is the root of $T$ and $p$ one of its children. In all other cases, let $M$ be the set of leaves descendant of $p$ in $T$. We claim that $M$ contains a leaf $x$ which is incident to a removed edge $xy$, with $y \notin M$. 
If $p$ is an affected leaf then this is true by definition. Otherwise, if $p$ is an affected node different from the root, assume by contradiction that all the removed edges in $M$ are of the form $uv$ with $u,v \in M$. In particular, this implies that $M$ is a module of $G$ strictly contained in a connected component. By Observation~\ref{obs:reduced-123}, it follows that $M$ is an independent set and hence contains no edges, a contradiction. 
Let $t$ be the least common ancestor of $x$ and $y$. The node $t$ is a parallel node which is an ancestor of $p$ and $q$ (observe that we may have $t = q$). Assume by contradiction that the path between $x$ and $t$ in $T$ contains a sequence of $2k+3$ consecutive non-affected nodes. The type of these nodes is alternatively series and parallel. So we can find a sequence $s_1, p_1\dots s_{k+1}, p_{k+1}$ of consecutive non-affected nodes with $s_i$ (resp. $p_i$) being the father of $p_i$ (resp. $s_{i+1}$) and with $s_i$'s being series nodes and the $p_i$'s being parallel node. Now each of the $s_i$'s (resp. $p_i$) has a non-affected leaf $a_i$ (resp. $b_i$) which is not a descendant of $p_i$ (resp. $s_{i+1}$). Observe that for every $i\in [1,k+1]$ the vertex set $\{b_i,a_i,x,y\}$ induces a $P_4$ in $G$. Thereby we found a set of $k+1$ $P_4$'s in $G$ pairwise intersecting on the edge $xy$. It follows that $G$ is not reduced by the Rule~\ref{rule:sunflower-del}: contradiction. This implies that the path between $p$ and $q$ contains at most $2k + 3$ non-affected nodes. 
\hfill $\diamond$ \\

Since there are at most $2k$ affected nodes and $2k$ affected leaves, $T$ contains at most $(4k-1)(2k+3) + 2k$ internal nodes.
As $G$ is reduced, Observation~\ref{obs:reduced-123} implies that each of these $O(k^2)$ nodes is attached to a set of at most $k+1$ leaves or a parallel node with $k+1$ children. It follows that $T$ contains at most $2k + (k+1)[(4k-1)(2k+3) + 2k] \leqslant 8k^3 + 20k^2 + 11k$ leaves, which correspond to the number of vertices of $G$. 

\medskip
We now conclude with the time complexity needed to compute the kernel. Since the application of Rule~\ref{rule:sunflower-del} decreases the value of the parameter (which is not changed by the other rules), Rule~\ref{rule:sunflower-del} is applied at most $k\leqslant n^2$ times. It then follows from Lemma~\ref{lem:gen-reduced} that a reduced instance can be computed in polynomial time.
 \end{proof}\\
 
The following corollary simply follows from the observation that the family of cographs is closed under complementation (since the complement
 of a $P_4$ is a $P_4$).
 
 \begin{corollary}
The parameterized cograph edge-completion problem admits a cubic vertex kernel.
\end{corollary}

\subsection{Cograph edge-edition}

The lines of the proof for the cubic kernel of the edge-edition problem are essentially the same as for the edge-deletion problem. But since edges can be added and deleted, the reduction Rule~\ref{rule:sunflower-del} has to be refined in order to avoid that a single edge addition breaks an arbitrary large set of $P_4$'s.

\begin{reduction} \label{rule:sunflower-edition}
If $\{x,y\}$ is a pair of vertices of $G$ that belongs to a set $S$ of  $t\geqslant k+1$ quadruples $P_i=\{x,y,a_i,b_i\}$ such that for $1\leqslant i\leqslant t$, every $P_i$ induces a $P_4$ and for any $1\leqslant i<j\leqslant t$, $P_i\cap P_j=\{x,y\}$, then change $E$ into $E\vartriangle \{xy\}$ and decrease $k$ by one.
\end{reduction}

As for reduction Rule~\ref{rule:sunflower-del}, it is clear that reduction Rule~\ref{rule:sunflower-edition} is safe and can be applied in polynomial time. The kernelization algorithm of cograph edge-edition consists of an exhaustive application of Rules~\ref{rule1}, \ref{rule2}, \ref{rule3} and \ref{rule:sunflower-edition}.

\begin{theorem}
\label{th:edition}
The parameterized cograph edge-edition problem has a cubic vertex kernel.
\end{theorem}

\begin{proof}
Let $G = (V,E)$ be a graph reduced under Rule~\ref{rule1}, Rule~\ref{rule2}, Rule~\ref{rule3} and Rule~\ref{rule:sunflower-edition} that can be turned into a cograph by editing at most $k$ edges. Let $H$ be the cograph obtained by an optimal edge-edition. The cotree of $H$ is denoted by $T$. Unlike in the edge-deletion problem, the root of $T$ is not necessary a parallel node. However it is still true that the root of $T$ is affected. Indeed, assume first that the root of $T$ is a series node. Then it is affected since otherwise $G$ would not be reduced under Rule~\ref{rule2}. Now, assume that the root is a non affected parallel node. This means that at most one of its children contains an affected leaf as descendant, and hence that $G$ is not reduced under Rule~\ref{rule1}: contradiction.

In the following we assume w.l.o.g. that the root of $T$ is a parallel node. We prove that Claim~\ref{claim:kernel-del} still holds in this case.
Let $p$ be an affected leaf or an affected node different from the root, and $q$ be the least affected ancestor of $p$. Observe that the result is trivially true if $q$ is the root of $T$ and $p$ one of its children. In all other cases, let $M$ be the set of leaves descendant of $p$ in $T$. As in the proof of Theorem~\ref{th:deletion}, there must exist an edited edge $xy$ with $x \in M, y \notin M$ (otherwise $M$ would be a module of $G$, i.e. an independent set by Observation~\ref{obs:reduced-123} and would thus not be edited by Observation~\ref{obs:module}). 

Now the proof follows the arguments of the proof of Theorem~\ref{th:deletion}, if one can find in $T$ a path of $2k+3$ consecutive non-affected nodes between $p$ and $q$, then $G$ is not reduced under Rule~\ref{rule:sunflower-edition}. Proving that $T$ contains $O(k^2)$ nodes and thereby $O(k^3)$ leaves.

\medskip
The fact that a reduced instance can be computed in polynomial time follows from Lemma~\ref{lem:gen-reduced} and the observation that Rule~\ref{rule:sunflower-edition} decreases the value of the parameter and requires polynomial time.
\end{proof}\\

\begin{figure}[ht]
\centerline{\includegraphics[scale=0.75]{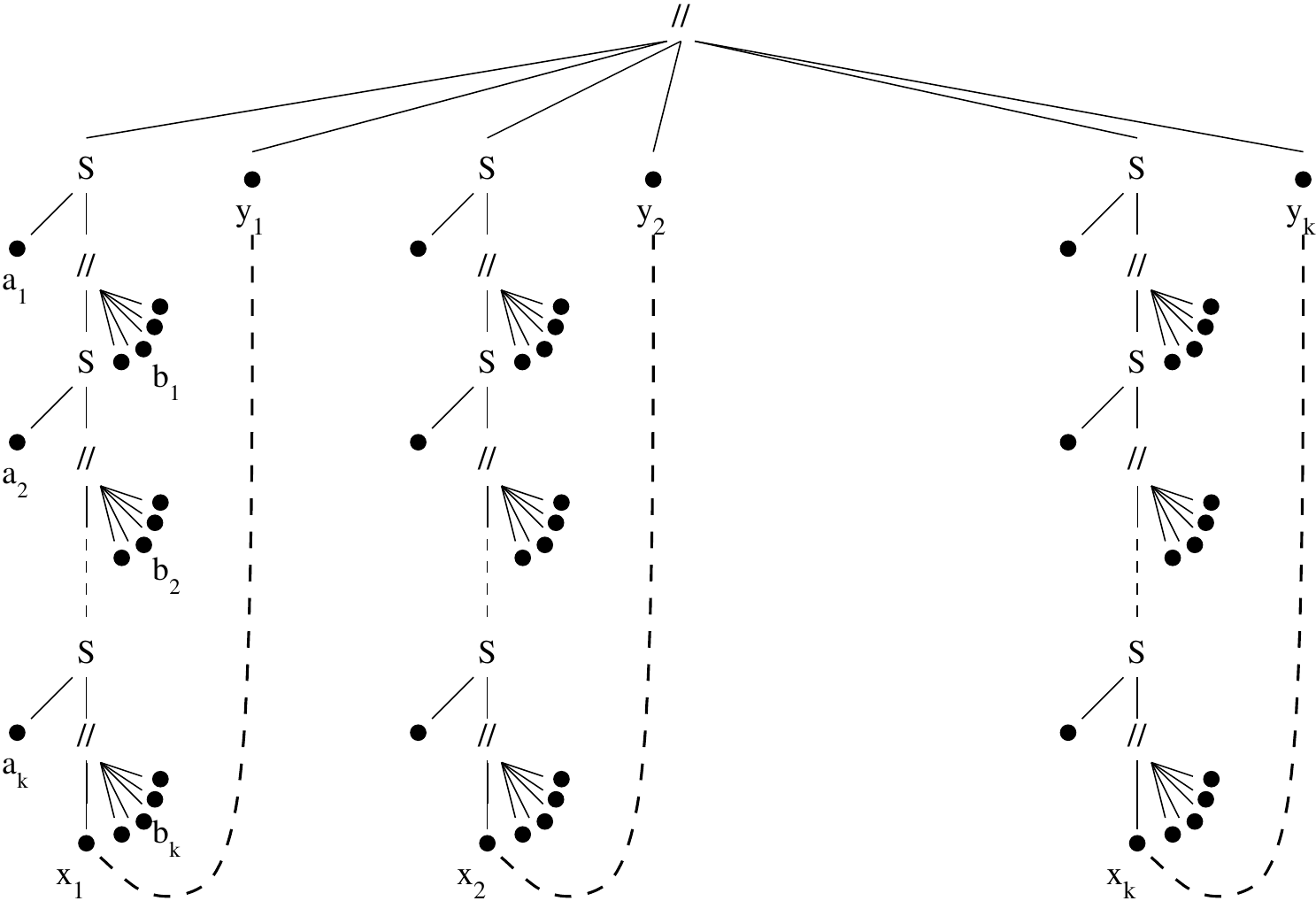}}
\caption{A reduced graph $G$ with $k(k+1)^2+k$ vertices for which $k$ edge deletions, namely the $x_iy_i$'s for $i\in[1,k]$, are required to obtain a cograph $H$. The cotree $T$ of $H$ is represented. Each parallel node of $T$  which is not the root has $k+2$ children, $k+1$ of which are leaves. The root of $T$ has $2k$ children.
\label{fig:lower-bound}}
\end{figure}

For the deletion (resp. edition) problem there exists a graph reduced under Rule~\ref{rule1}, Rule~\ref{rule2}, Rule~\ref{rule3} and Rule~\ref{rule:sunflower-del} (resp. Rule~\ref{rule:sunflower-edition}) that achieves the cubic bound (see Figure~\ref{fig:lower-bound}).

\section{Kernel lower bounds for $P_l$-free edge-deletion problems}
\label{sec:lower}

\newcommand{\notone}{\textsc{Not-1-in-3-sat}}

In \cite{KW09}, Kratsch and Wahlstr\"om show that the \notone{} problem has no polynomial kernelization under a complexity-theoretic assumption ($NP \nsubseteq coNP / poly$).
We observe that their argument still applies to a graph restriction of \notone{} where the constraints arise from the triangles of an input graph. 

\subsection{A graphic version of the {\sc\notone{}} problem}

For a graph $G=(V,E)$, an edge-bicoloring is a function $B: E \rightarrow \{0,1\}$. A \emph{partial edge-bicoloring} of $G$ is an edge-bicoloring of a subset of edges of $E$. An edge colored $1$ (resp. $0$) is called a $1$-edge (resp. $0$-edge). We say that the edge-bicoloring $B'$ \emph{extends} a partial edge-bicoloring $B$ if for every $e\in E$ colored by $B$, then $B(e)=B'(e)$.
The weight of an edge-bicoloring is the number $\omega(B)$ of $1$-edges. An edge-bicoloring is \emph{valid} if  every triangle of $G$ contains either zero, two or three $1$-edges. We consider the following problem:

\medskip
\noindent
\textsc{Not-1-in-3-edge-triangle}\\
\textbf{Input:} An undirected graph $G=(V,E)$ and a partial edge-bicoloring $B: E \rightarrow \{0,1\}$.\\
\textbf{Parameter:} An integer $k\in\mathbb{N}$.\\
\textbf{Question:} Can we extend $B$ to a valid edge-bicoloring $B'$ of weight at most $k$?

\begin{proposition} \label{prop1} 
 \gbc{} is NP-complete and or-compositional.
\end{proposition}

\begin{proof}
The NP-hardness follows from a reduction from \textsc{Vertex Cover}. Let $(G,k)$ be an instance of \textsc{Vertex Cover}~\cite{GJ79}, where $G = (V,E)$. We create an instance $(G',B,k')$ of \gbc{} as follows. The graph $G'$ is obtained from $G$ by adding a dominating vertex $q$, the partial edge-bicoloring $B$ is such that $B(e) = 1$ for every $e \in E$, and we let $k' = |E|+k$. As the triangles of $G$ are monochromatic, the constraints to obtain a valid extension of $B$ are carried by the triangles of the form $quv$ with $uv\in E$. It is easy to observe that $(G',B,k')$ has a valid edge-bicoloring extension of weight $k'$ iff $G$ has a vertex cover of size $k$. As \gbc{} clearly belongs to NP, the NP-completeness follows.

\medskip
We now show that \gbc{} is or-compositional. The proof closely follows the proof of \cite{KW09} for \notone{}. We first need the following result:

\begin{claim}\label{lemma1} 
Given  an instance $(G,B,k)$ of \gbc{}, and two positive integers $r$ and $k'$ such that $k' \geq k+r$, we can compute in polynomial time an equivalent instance $(G',B',k')$ of \gbc{} such that $\omega(B') = \omega(B)+r$.
\end{claim}

\emph{Proof.} To build $G'$, we first add to $G$ a set $F$ of $r$ new isolated edges $e_1\dots e_r$ such that $B'(e_i)=1$ for all $i\in[r]$. Then we add to the resulting graph $k'-(k+r)$ gadgets as follows: let $e_j=u_jv_j$ (with $j\in[k'-(k+r)]$) be an arbitrary $1$-edge of $G$; add the triangles $u_jv_jx_j,v_jx_jy_j$ with $B'(v_jy_j)=B'(x_jy_j)=0$. The edges $e_j$'s are not necessarily distinct. Observe that in any valid edge-bicoloring of $G'$ extending $B'$, the edge $v_jx_j$ (for every $j\in[k'-(k+r)]$) is a $0$-edge while the edge $u_jx_j$ is a $1$-edge. It follows that $(G,B,k)$ is a positive instance if and only if $(G',B',k')$ is a positive instance as the set $F$ increases the weight by $r$ and the added triangles by $k'-(k+r)$.
\hfill $\diamond$ \\

Consider a sequence $(G_1,B_1,k)\dots (G_t,B_t,k)$ of instances of \gbc{}. 
We denote by $E_1(j)$ the set of $1$-edges of $(G_j,B_j,k)$. 
By Claim \ref{lemma1}, we can assume w.l.o.g. that $|E_1(j)|=s \leqslant k$, for $1\leqslant j\leqslant t$. We can also assume that $t\leqslant 3^k$ since otherwise an exact branching algorithm could solve the problem. Moreover, for the sake of the construction, we assume $t=2^l$ (duplicating some instance $(G_i, B_i, k)$ if necessary).

Intuitively, the graph $G$ of the composed instance $(G,B,k')$ is built on the disjoint union of the $G_j$'s, $1\leqslant j\leqslant t$.  Then, as a selection gadget, we add a "tree-like graph" $T$ connecting a "root edge" $r$ to edges $e_j$ for $j = 1,...,t$. Finally, for every $1\leqslant j\leqslant t$, the $1$-edges of the graph $G_j$ are connected via a propagation gadget to the edge $e_j$ in $T$. The root edge is the unique $1$-edge of $G$. The copies of the $G_j$'s inherit the $0$-edges of the $G_j$'s. The idea is that the selection gadget guarantees that at least one of the $e_j$'s edge gets colored $1$. Then the propagation gadgets attached to that edge $e_j$ transmit color $1$ to the copies of every $1$-edge of $G_j$.

Formally, we do the following: (i) we start with a complete binary tree $T_0$ with $t$ leaves; (ii) to each node $u$ of $T_0$, we associate an edge $e_u$ in $T$ as follows: if $u$ is associated to the edge $xy$ and if $u$ has two children $v,v'$, we create a new vertex $z$ and we let $e_v = xz, e_{v'} = yz$. The leaves of $T_0$ are then associated to edges $e_1,...,e_t$. Now, for every $1\leqslant j\leqslant t$, the propagation gadget $S_j$ consists of vertex-disjoint graphs $S_{j,e}$ for every edge $e$ of $E_1(j)$. If $e = uv$ and $e_j = xy$, then $S_{j,e}$ consists of four triangles $uva, vab, abx, bxy$, with edges $ua,vb,ax,by$ colored $0$ by $B$ (the other edges remain uncolored). Again the unique $1$-edge of $B$ is the root edge of $T$, in particular the edges of the $E_1(j)$ are uncolored by $B$. However, the $0$-edge sets of the $G_j$'s are inherited by $B$. (see Figure~\ref{fig:not-1-in-3-triangle})

\begin{figure}[ht]
\centerline{\includegraphics[scale=0.75]{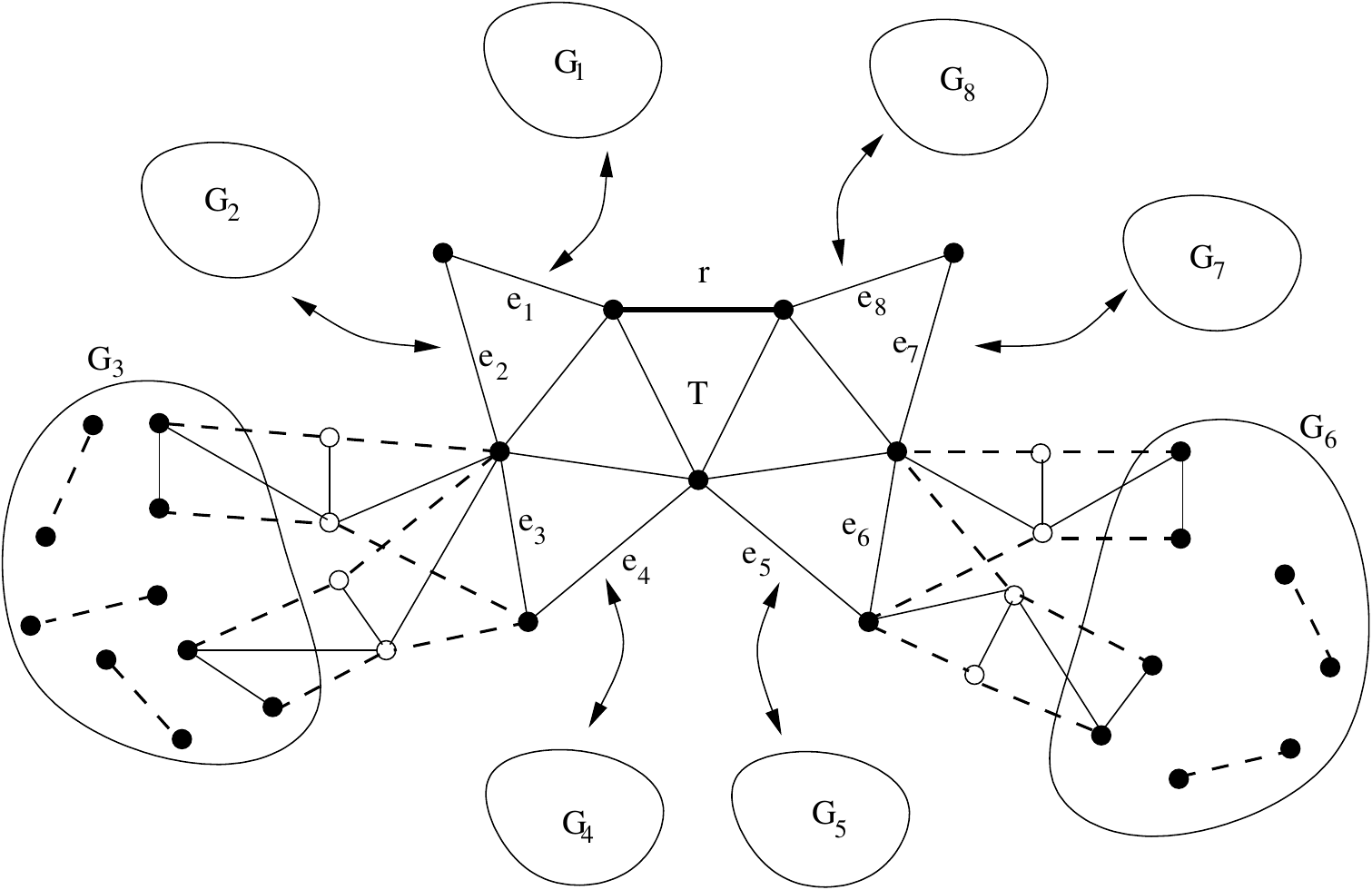}}
\caption{The instance $(G,B,k')$ built from a sequence $(G_1,B_1,k),\dots,(G_t,B_t,k)$ with $t=2^3$. The unique $1$-edge is $r$. Every "leaf edge" $e_j$ of $T$ is linked to the copies of the $1$-edges of $(G_j,B_j,k)$ via the propagation gadget. The $0$-edges are depicted as dotted lines: they either belong to a propagation gadget or correspond to a $0$-edge of some $(G_j,B_j,k)$.
\label{fig:not-1-in-3-triangle}}
\end{figure}

Observe first that every valid edge-bicoloring extending $B$ has to assign color $1$ to at least one edge $e_j$, for $1\leqslant j\leqslant t$. Then the edges of $E_1(j)$ and the $3s$ non $0$-edges of $S_j$ are also assigned color $1$. It follows that if we choose $k'=k+3s+l$, then $(G,B,k')$ is a positive instance if and only if there exists $1\leqslant j\leqslant t$ such that $(G_j,B_j,k)$ is a positive instance.
 \end{proof}\\

The following corollary follows from Theorem~\ref{th:or-comp}:

\begin{corollary}
The \gbc{} problem does not admit a polynomial kernel unless $NP\subseteq coNP/poly$.
\end{corollary}

The problem \tgbc{} is the restriction of \gbc{} where the input graph $G$ is 3-colorable. The hardness results obtained for \gbc{} carry over to this restriction:

\begin{lemma} \label{prop2} 
The \tgbc{} problem  does not admit a polynomial kernel unless $\hyp$.
\end{lemma}

\begin{proof}
The proof uses Theorem~\ref{th:ppt}, that is we provide a polynomial parameter-preserving transformation from \gbc{} to \tgbc{}. By Proposition \ref{prop1}, \gbc{} is NP-complete. Observe that \tgbc{} clearly belongs to NP.

Let $(G,B,k)$ be an instance of \gbc{}. We build an instance $(G',B',6k)$ of \tgbc{} in the following way. Suppose that $G = (V,E)$, then $G'$ has vertex set $V' = \{ v_1,v_2,v_3 : v \in V \}$, and has edge set $E' = \{ u_1 v_2, u_1 v_3, u_2 v_3 : u = v \text{ or } u v \in E \}$. The partial edge-bicoloring $B'$ is defined as follows: $B'(u_iu_j)=0$ for $1\leqslant i<j\leqslant 3$; if the edge $uv$ of $G$ is colored, then $B'(u_iv_j)=B(uv)$ for $1\leqslant i,j\leqslant 3$; the other edges of $G'$ are uncolored.

Observe that every valid edge-bicoloring extending $B'$ assigns the same color to the six edges of $G'$ associated with an edge $uv$ of $G$: indeed, given $u_i v_j, u_k v_l$ $1\leqslant i,j,k,l\leqslant 3$, if $i= j$ this holds since $B'(v_k v_l) = 0$, if $k = l$ this holds since $B'(u_i u_j) = 0$, and otherwise this follows by transitivity. It is then easy to see that solutions of $(G,B,k)$ and solutions of $(G',B',6k)$ are in one-to-one correspondence. 
\end{proof}

\subsection{Negative results for $\Gamma$-free edge deletion problems}

In this section, we show that unless $NP \subseteq coNP / poly$, the \textsc{$C_l$-free edge-deletion} and the \textsc{$P_l$-free edge-deletion} problems have no polynomial kernel for large enough $l\in\mathbb{N}$. To that aim, we provide polynomial time and parameter transformations from \tgbc{} to the \textsc{Annotated $C_l$-free edge-deletion} problem and to the \textsc{Annotated $P_l$-free edge-deletion} problem. For a graph $\Gamma$, the \textsc{Annotated $\Gamma$-free edge-deletion} problem is defined as follows:\\

\noindent
\textsc{Annotated $\Gamma$-free edge-deletion}\\
\textbf{Input:} An undirected graph $G=(V,E)$ and a subset $S$ of vertices.\\
\textbf{Parameter:} An integer $k\in\mathbb{N}$.\\
\textbf{Question:} Is there a subset $F \subseteq E\cap(S\times S)$ such that $H=(V,E\setminus F)$ is $\Gamma$-free?\\

Observe that the \textsc{Annotated $\Gamma$-free edge-deletion} problem reduces to the (unannotated) \textsc{$\Gamma$-free edge-deletion} problem whenever $\Gamma$ is closed under twin addition: it suffices to add for every vertex $v\in V\setminus S$ a set of $k+1$ twin vertices. Clearly this transformation also preserves the parameter.  \\

Observe also that we can restrict the \tgbc{} problem to instances $(G,B,k)$ not containing any $0$-edge (\emph{i.e.} $B(e) = 1$ whenever it is defined). The reason is that any uncolored edge $e=uw$ of $G$ can be forced to be assigned color $0$ in every valid edge-bicoloring extending $B$ by adding to $G$ $k+1$ new vertices $v_1,\dots,v_{k}$ such that $uv_iw$, $1\leqslant i\leqslant k$, is an uncolored triangle. Clearly if $e$ is a $1$-edge of an edge-bicoloring $B'$ extending $B$, $B'$ needs at least $k+1$ $1$-edges to be valid: $e$ plus one edge per triangle. The same argument was used in~\cite{KW09} for the \notone{} problem.

\newpage

\begin{theorem} \label{prop3} 
The \textsc{$C_l$-free edge-deletion} problem has no polynomial kernel for any $l \geqslant 12$, unless $NP \subseteq coNP / poly$.
\end{theorem}

\begin{proof}
We describe a polynomial time and parameter transformation from the restriction of \tgbc{} without $0$-edges to \textsc{Annotated $C_l$-free edge-deletion}. The statement then follows from Theorem~\ref{th:ppt} and the fact that \textsc{Annotated $C_l$-free edge-deletion} reduces to \textsc{$C_l$-free edge-deletion}.

Let $(G,B,k)$ be an instance of the \tgbc{} problem, where $V_1,V_2,V_3$ are disjoint independent sets of $G=(V,E)$. The construction of the instance $(H,S,k')$ of \textsc{Annotated $C_l$-free edge-deletion} works as follows. First the sets $V_1$, $V_2$ and $V_3$ are turned into cliques and the $1$-edges of $G$ are removed. In addition to $V$, the graph $H$ contains a set $U$ of new vertices. For each pair $t=(e,v)$ with $e=uw$ an edge of $G$ and $v$ a vertex of $G$, such that $\{u,v,w\}$ induces a triangle in $G$, we create a path $P_t$ of length $l-1$ between $u$ and $w$ in $H$ (the internal vertices of $P_t$ are added to $U$). Notice that each triangle of $G$ generates three such paths in $H$. It remains to add some \emph{safety} edges incident to the vertices of $U$. Every two vertices $x$ and $y$ of $U$ that do not belong to the same path are made adjacent.
In every path $P_t$, we select an internal vertex $c_t$, called its \emph{centre}, at distance $(l-1)/2$ from $u$. Every centre vertex $c_t$ is made adjacent to $V\setminus\{u,v,w\}$. We denote by $H=(V_H,E_H)$ the resulting graph. 
To complete the description of $(H,S,k')$ we set $S=V$ and the parameter $k'=k-k_1$ where $k_1$ is the number of $1$-edges of $(G,B,k)$.

\begin{figure}[ht]
\centerline{\includegraphics[scale=0.75]{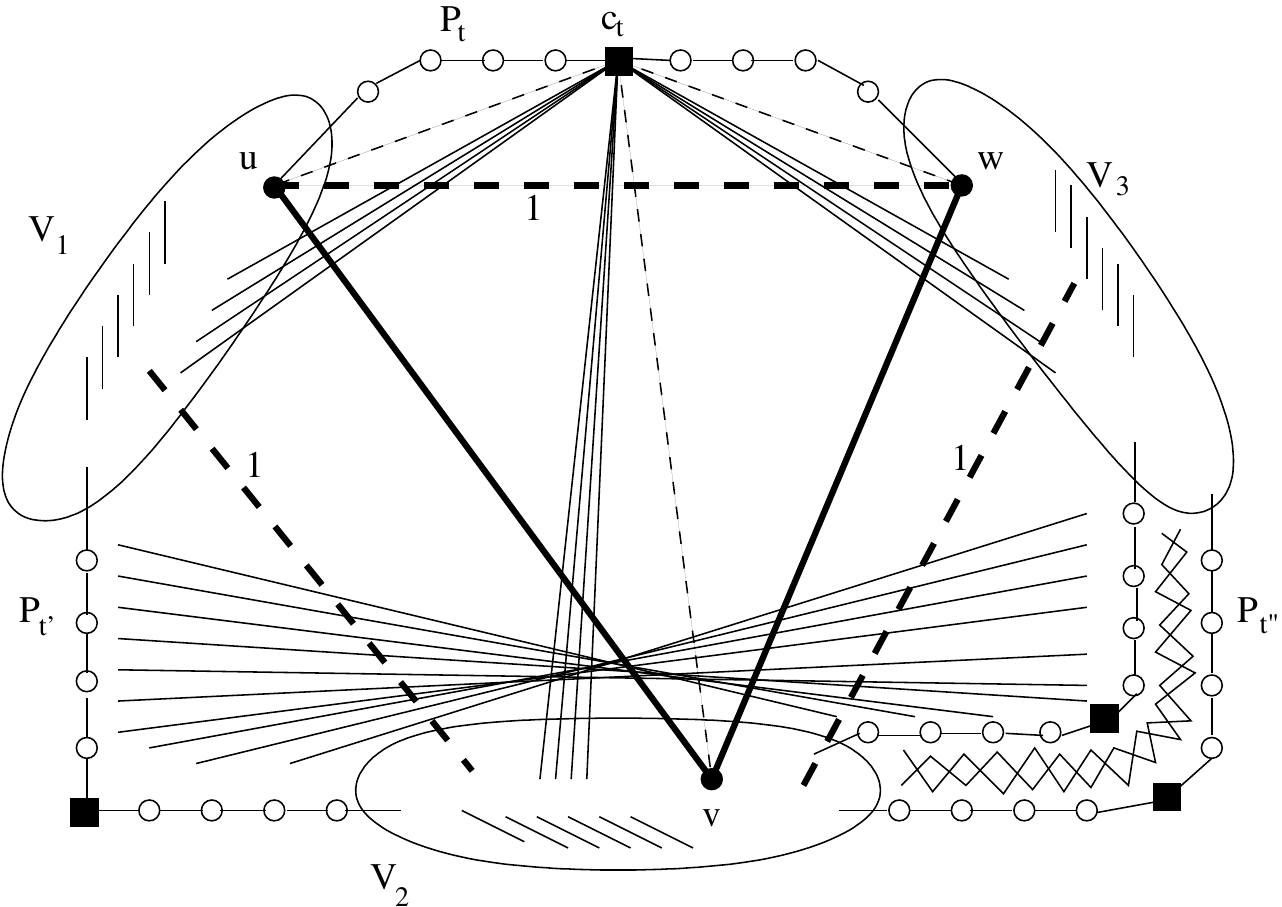}}
\caption{The graph $H=(V_H,E_H)$ built from an instance $(G,B,k)$ of the \tgbc{} problem for $l=12$. The white and the square vertices form the set $U$ of new vertices. The independent sets $V_1$, $V_2$ and $V_3$ of $G$ are turned into cliques. The thick dotted edges are the removed $1$-edges of $(G,B,k)$. The non $1$-edges of $(G,B,k)$ are preserved in $H$. 
\label{fig:cycle}}
\end{figure}

\begin{claim}
\label{claim:nokernel-cl}
A subset of vertices $C\in V_H$ induces a cycle of length $l$ iff $G$ contains a triangle $uvw$,  
with $e=uw$ a $1$-edge and $uv$, $vw$ uncolored edges, such that $C=P_{t}\cup\{v\}$ with $t=(e,v)$.
\end{claim}

\emph{Proof.} By construction, if $G$ contains a triangle $uvw$ with a unique $1$-edge $e=uw$, then $C=P_t\cup\{v\}$ (with $t=(e,v)$) induces a cycle of length $l$ in $H$ (keep in mind that the $1$-edges of $G$ are removed from $H$).
Let $C$ be an induced $C_l$ in $H$. Observe that as $V_1$, $V_2$ and $V_3$ are turned into cliques, $|C\cap V|\leqslant 6$. Thereby $C$ intersects the vertex set $U$. We now argue that there exists a path $P_t$, with $t=(e,v)$ and $e=uw$, containing the vertices of $C\cap U$. Otherwise, since every pair of vertices of $U$ belonging to two distinct paths $P_t$ and $P_{t'}$ (with $t\neq t'$) are adjacent, we would have $|C\cap U|\leqslant 4$ and thus $|C|\leqslant 4+6<l$. It follows that $u$ or $w$ belongs to $C$. We prove that they both belong to $C$. Assume $w\notin C$, then $C$ uses a safety edge incident to the centre vertex $c_t$ and half of the internal vertices of $P_t$ does not belong to $C$. Thereby $|C|\leqslant 6+(l-3)/2+1$: contradiction with the hypothesis $l\geqslant 12$. Finally as $P_t$ contains $l-1$ vertices, $C$ contains an extra vertex and $uw$ are not adjacent. As $c_t$ is adjacent to every vertex of $V$ except $u$, $v$ and $w$, we have that $C=P_t\cap\{v\}$ as announced and $uv,wv\in E_H$. Now the existence of $P_t$ witnesses the existence of the triangle $uvw$ in $G$. As $uv,wv\in E_H$ and $uw\notin E_H$, $uw$ is the only $1$-edge of the triangle $uvw$.
\hfill $\diamond$ \\

We now argue for the correctness of the transformation. Suppose that there exists a set $F$ of allowed edges of size at most $k'$ such that $H'=(V_H,E_H\setminus F)$ is $C_l$-free. Define the edge-bicoloring $B'$ of $E$ as follows: $B'(e) = 1$ if $e \in F$, $B'(e) = 0$ otherwise. As by assumption $B$ does not assign color $0$ to any edge, $B'$ extends $B$ and has weight at most $|F|+k_1\leqslant k'+k_1=k$.
Besides, $B'$ is a valid edge-bicoloring of $G$. Let $t=(e,v)$ with $e=uw$ be a pair such that $\{u,v,w\}$ induces a triangle in $G$. If we had $B(uw) = 1$, $B'(uv)=B'(vw)=0$, we would obtain that $P_t\cup\{v\}$ induces a $C_l$ in $H'$, impossible. Conversely, suppose that $B'$ is valid edge-bicoloring of weight at most $k$ of $G$ which extends $B$. Let $F\subseteq E$ be the set of edges such that $B'(e)=1$ but are uncolored by $B$. By construction $F$ is a set of allowed edges of $H$ of size at most $k-k_1$. Since $B'$ is a valid edge-bicoloring of $G$, Claim~\ref{claim:nokernel-cl} implies that $H'=(V_H,E_H\setminus F)$ is $C_l$-free. 
 \end{proof}\\
 
A slight modification of the above construction yields the following:

\begin{theorem} \label{prop4} 
The \textsc{$P_l$-free edge-deletion} problem has no polynomial kernel for any $l \geqslant 13$, unless $NP \subseteq coNP / poly$.
\end{theorem}

\begin{proof}
Let $(G,B,k)$ be an instance of the \tgbc{} problem not containing any $0$-edge and such that $V_1,V_2,V_3$ are disjoint independent sets of $G=(V,E)$. We modify the construction given in Theorem \ref{prop3} to obtain an instance $(H,S,k')$ of \textsc{Annotated $P_l$-free edge-deletion} problem. The vertex set $V_H$ of $H$ consists of the union of $V$ and a set $U$ of new vertices. The sets $V_1$, $V_2$ and $V_3$ are again turned into cliques and the $1$-edges of $E$ are not duplicated in $E_H$. But for each pair $t=(e,v)$, with $e=uw\in E$ and $v\in V$ such that $\{u,v,w\}$ is a triangle of $G$, the associated gadget $Q_t$ is no longer a path. Instead, $Q_t$ consist of two paths $Q_t^u$ and $Q_t^w$: $Q_t^u$ is a path of length $(l-1)/3$ containing $u$ as extremity and $Q_t^w$ is a path of length $2(l-1)/3$ containing $w$ as extremity. The vertices of $Q_t\setminus\{u,w\}$ are added to $U$. As before for every $t\neq t'$ we add all the edges between vertices of $Q_t$ and $Q_{t'}$. The centre vertex $c_t$ of $Q_t$ is the vertex of $Q_t^w$ at distance $(l-1)/3$ from $w$. The centre vertex is made adjacent to every vertex of $V$ except $u$, $v$ and $w$. To complete the description of $(H,S,k')$ we set $S=V$ and $k'=k-k_1$ where $k_1$ is the number of $1$-edges of $(G,B,k)$.

The correctness proof of the construction follows the same lines than the proof of Proposition~\ref{prop3}. It now relies on the following claim that characterizes the possible induced $P_l$'s.

\begin{claim}
\label{claim:nokernel-pl}
A subset of vertices $Q\in V_H$ induces a path of length $l$ iff $G$ contains a triangle $uvw$,  
with $e=uw$ a $1$-edge and $uv$, $vw$ uncolored edges, such that $Q=Q_{t}\cup\{v\}$ with $t=(e,v)$.
\end{claim}

\emph{Proof.} By construction, if $G$ contains a triangle $uvw$ with a unique $1$-edge $e=uw$, then $Q=Q_t\cup\{v\}$ (with $t=(e,v)$) induces a path of length $l$ in $H$ (keep in mind that the $1$-edges of $G$ are removed from $H$).
Let $Q$ be an induced $P_l$ in $H$. As in the proof of Claim~\ref{claim:nokernel-cl}, observe that $|Q\cap V|\leqslant 6$ and thereby $Q$ intersects the vertex set $U$ and that there exists a unique pair $t=(e,v)$ with $e=uw$ such that $Q_t$ contains $Q\cap U$. By the choice of the length of $Q_t^u$ and $Q_t^w$, $Q\cap U$ intersects both $Q_t^u$ and $Q_t^w$. It follows that $u$ and $w$ belongs to $Q$. Assume that $Q$ uses an edge $xc_t$ such that $x\notin Q_t$. Then half of the vertices of $Q_t^w$ does not belong to $Q$, which would contradict the hypothesis $l\geqslant 13$. Finally as by construction $Q_t$ contains $l-1$ vertices, we need at least one extra vertex from $V$. Since $c_t$ is adjacent to all vertices of $V$ except $u$, $v$ and $v$, that extra vertex can only be $v$. Moreover the chord $uw$ cannot exist in $H$, meaning that $uw$ is a $1$-edge of $(G,B,k)$.
\end{proof}
 
 \section{Conclusion}
 
In this paper we have shown that the \textsc{parameterized cograph edge modification} problems admit vertex cubic kernels. Moreover, we provide evidence that the \textsc{$C_l$-free edge-deletion} and the \textsc{$P_l$-free edge-deletion} problems do not admit polynomial kernels for large enough $l$ (under a complexity-theoretic assumption~\cite{BDFH08}). These problems were left open by Kratsch and Wahlstr\"om in~\cite{KW09}. The value of $l$ being respectively (at least) $12$ and $13$, one remaining question is thus to determine whether the \textsc{$C_p$-free edge-deletion} and the \textsc{$P_p$-free edge-deletion} problems admit polynomial kernels for $4 \leqslant p < 12$ in the former case, and $4 < p < 13$ in the latter. 

\bibliographystyle{abbrv}
\bibliography{cograph}

\begin{thebibliography}{10}

\bibitem{BDFH08}
H.~Bodlaender, R.~Downey, M.~Fellows, and D.~Hermelin.
\newblock On problems without polynomial kernels.
\newblock In {\em ICALP}, number 5125 in LNCS, pages 563--574, 2008.

\bibitem{BTY09}
H.~L. Bodlaender, S.~Thomass{\'e}, and A.~Yeo.
\newblock Kernel bounds for disjoint cycles and disjoint paths.
\newblock In {\em ESA}, volume 5757 of {\em LNCS}, pages 635--646, 2009.

\bibitem{BLS99}
A.~Brandst{\" a}dt, V.~B. Le, and J.~P. Spinrad.
\newblock {\em Graph Classes: {A} Survey}.
\newblock SIAM Monographs on Discrete Mathematics and Applications. 1999.

\bibitem{Cai96}
L.~Cai.
\newblock Fixed-parameter tractability of graph modification problems for
  hereditary properties.
\newblock {\em Information Processing Letters}, 58(4):171--176, 1996.

\bibitem{DF99}
R.~Downey and M.~Fellows.
\newblock {\em Parameterized complexity}.
\newblock Springer, 1999.

\bibitem{EC88}
E.~S. El-Mallah and C.~Colbourn.
\newblock The complexity of some edge deletion problems.
\newblock {\em IEEE Transactions on Circuits and Systems}, 35(3):354--362,
  1988.

\bibitem{FLR07}
M.~Fellows, M.~Langston, F.~Rosamond, and P.~Shaw.
\newblock Efficient parameterized preprocessing for cluster editing.
\newblock In {\em Fundamentals of Computation Theory, 16th International
  Symposium, (FCT)}, number 4639 in LNCS, pages 312--321, 2007.

\bibitem{FG06}
J.~Flum and M.~Grohe.
\newblock {\em Parameterized complexity theorey}.
\newblock Texts in Theoretical Computer Science. Springer, 2006.

\bibitem{FS08}
L.~Fortnow and R.~Santhanam.
\newblock Infeasibility of instance compression and succinct {PCP}s for {NP}.
\newblock In {\em STOC}, pages 133--142, 2008.

\bibitem{GJ79}
M.~Garey and S.~Johnson.
\newblock {\em Computers and intractability: a guide to the theory of
  NP-completeness}.
\newblock Freeman, 1978.

\bibitem{GKS95}
M.~Golumbic, H.~Kaplan, and R.~Shamir.
\newblock Graph sandwich problems.
\newblock {\em Journal of Algorithms}, 19:449--473, 1995.

\bibitem{HP10}
M.~Habib and C.~Paul.
\newblock A survey on algorithmic aspects of modular decomposition.
\newblock {\em Computer Science Review}, 4(1):41--59, 2010.

\bibitem{HPTV07}
P.~Heggernes, C.~Paul, J.~A. Telle, and Y.~Villanger.
\newblock Interval completion with few edges.
\newblock In {\em STOC}, pages 374--381, 2007.

\bibitem{KS07}
C.~Kenyon-Mathieu and W.~Schudy.
\newblock How to rank with few errors.
\newblock In {\em Annual ACM Symposium on Theory of Computing (STOC)}, pages
  95--103, 2007.

\bibitem{KW09}
S.~Kratsch and M.~Wahlstr{\"o}m.
\newblock Two edge modification problems without polynomial kernels.
\newblock In {\em {IWPEC}}, volume 5917 of {\em LNCS}, pages 264--275, 2009.

\bibitem{NSS01}
A.~Natanzon, R.~Shamir, and R.~Sharan.
\newblock Complexity classification of some edge modification problems.
\newblock {\em Discrete Applied Mathematics}, 113(1):109 -- 128, 2001.

\bibitem{Nie06}
R.~Niedermeier.
\newblock {\em Invitation to fixed parameter algorithms}, volume~31 of {\em
  Oxford Lectures Series in Mathematics and its Applications}.
\newblock Oxford University Press, 2006.

\bibitem{NR00}
R.~Niedermeier and P.~Rossmanith.
\newblock A general method to speed up fixed-parameter-tractable algorithms.
\newblock {\em Information Processing Letters}, 73(3-4):125--129, 2000.

\bibitem{Ros72}
D.~Rose.
\newblock A graph-theoretic study of the numerical solution of sparse positive
  systems of linear equations.
\newblock {\em Graph Theory and Computing}, pages 183--217, 1972.

\bibitem{SST04}
R.~Shamir, R.~Sharan, and D.~Tsur.
\newblock Cluster graph modification problems.
\newblock {\em Discrete Applied Mathematics}, 144(1-2):173--182, 2004.

\bibitem{TY84}
R.~Tarjan and M.~Yannakakis.
\newblock Simple linear-time algorithms to test chordality of graphs, test
  acyclicity of hypergraphs, and selectively reduce acyclic hypergraphs.
\newblock {\em SIAM journal of Computing}, 13:566--579, 1984.

\bibitem{ZW08}
A.~van Zuylen and D.~Williamson.
\newblock Deterministic algorithms for rank aggragation and other ranking and
  clustering problems.
\newblock In {\em WAOA}, volume 4927 of {\em LNCS}, pages 260--273, 2008.

\end{thebibliography}

\end{document}